\documentclass[preprint,aip,superscriptaddress,jap,onecolumn]{revtex4-1}
\usepackage{graphicx}
\usepackage{amsmath}
\usepackage[english]{babel}
\usepackage{color}

\begin{document}

\title{Controlled emission and coupling of small-size YAG:Ce$^{3+}$ nanocrystals to gold nanowire}
\author{Muhammad Danang Birowosuto}
\affiliation{Physics Research Center, The Indonesian Institute of Sciences, Puspitek, Serpong, Banten 15314, Indonesia}
\affiliation{CINTRA UMI CNRS/NTU/THALES 3288, Research Techno Plaza, 50 Nanyang Drive, Border X Block, Level 6, Singapore 637553, Singapore}
\author{Isnaeni}
\affiliation{Physics Research Center, The Indonesian Institute of Sciences, Puspitek, Serpong, Banten 15314, Indonesia}
\author{Celso de Mello Donega}
\affiliation{Debye Institute, Condensed Matter and Interfaces, Utrecht University, P.O. Box 80 000, Utrecht, The Netherlands}
\author{Andries Meijerink}
\affiliation{Debye Institute, Condensed Matter and Interfaces, Utrecht University, P.O. Box 80 000, Utrecht, The Netherlands}
\date{\today}

\begin{abstract}
We report a controlled emission of \textcolor{red}{Ce$^{3+}$} ions inside single Yttrium Aluminum Garnet $Y_{3}Al_{5}O_{12}$ (YAG:Ce$^{3+}$) nanocrystals with a diameter of 22 $\pm$ 10 nm as a result of a coupling of a surface plasmon mode propagating along single gold nanowire (NW). From the PL images, the intensity for single YAG:Ce$^{3+}$ nanocrystals in the proximity of the single gold NW increases by a factor of two in comparison with that without the NW. Also, we observe a maximum of 3.8-fold emission rate enhancements for the single nanocrystal close to the single gold NW. The emission rate enhancements of YAG:Ce$^{3+}$ nanocrystals are two folds the enhancements of 100-nm fluorescent nanospheres. This value is in agreement with the calculation of a combination from the analytical scattering model and boundary element method (BEM). We also confirm that the small size light sources are more efficient for the emission rate enhancements. Thus, the controlled emission of small YAG:Ce$^{3+}$ nanocrystals with the perfect photostabilities will pave the way for the ultimate efficient nanoscale light sources.    
\end{abstract}

\maketitle

\section{Introduction}

Lanthanide-doped nanocrystals are especially promising since their light is emitted from a single atom which is shielded from the environment. As such, these new sources are immune to the two major disadvantages of today’s emitters: the temporary loss of fluorescence (blinking) which is a common feature of quantum dots \cite{Nirmal1996} and permanent loss of fluorescence (photobleaching) which is the major disadvantage of molecular emitters \cite{Fleur1998}. The reliable emission and perfect photostability of lanthanide-doped nanocrystals promise the full control of light-matter interaction at the nanoscale. As a dopant, Cerium (Ce$^{3+}$) has strong advantages compared to other lanthanides \cite{Birowosuto2009}. Ce$^{3+}$-doped nanocrystals have high quantum efficiency due to 5d-4f dipole Ce$^{3+}$ allowed transition. Because of the large oscillator strength of the transition, they are extremely bright and fast.

A particularly interesting Ce$^{3+}$-doped nanocrystal is Ce$^{3+}$-doped Yttrium Aluminum Garnet $Y_{3}Al_{5}O_{12}$ (YAG:Ce$^{3+}$). In bulk, YAG:Ce$^{3+}$ is a well-known inorganic compound which has excellent chemical, physical and optical properties \cite{Wong1984}. This crystal emits yellow light on excitation with blue light and it is widely applied for solid-state lighting, lasers and display devices \cite{Nikl2007}. It is reported that YAG:Ce$^{3+}$ nanocrystals perform better than larger particles for white LED application \cite{Pan2004} while they also exhibit no blinking and no bleaching \cite{Wuister2004} while Ce$^{3+}$ emission lifetime in YAG of 60 ns is relatively four times slower with that of the fastest of 15 ns in Ce$^{3+}$ with different hosts \cite{Birowosuto2009}. Therefore, we still can optimize the radiative recombination rate of single YAG:Ce$^{3+}$ nanocrystals. 

Here we control \textcolor{red}{the 5d-4f emission of Ce$^{3+}$} in single YAG:Ce$^{3+}$ nanocrystal by the manipulation of the local density of photonic states (LDOS) \cite{Sprik1996}. Controlling the spontaneous emission has been demonstrated for large varieties of the systems such as planar mirrors \cite{Drexhage1970}, nanowire \cite{Birowosuto2014a}, photonic crystals \cite{Lodahl2004,Birowosuto2012,Birowosuto2012b,Birowosuto2014}, and localized plasmons \cite{DEChang2006,Murai2013}. Unlike other systems, to obtain the enhancement and the efficient coupling to the specific spatial mode at the same time, we use propagating plasmons on metallic nanowires (NWs) \cite{Akimov2007,Kolesov2009,AlexHuck2011,Frimmer2011}. Using such method, we may enhance the emissive process into one particular propagating plasmon mode which can be further transferred into a photonic mode of an optical waveguide with high efficiency \cite{Pyayt2008,AlexHuck2013a,AlexHuck2013}. In the previous experiments, the coupling of individual single light sources to propagating surface plasmon modes on individual metallic NWs has been demonstrated for CdSe quantum dots \cite{Akimov2007}, fluorescent dyed spheres \cite{Frimmer2011}, and for NV centers in nanocrystal diamonds \cite{Kolesov2009,AlexHuck2011}, where most emission rate enhancements were in the range of 1.7-2.5 folds. 
\textcolor{red}{They are many reports on the emission rate enhancements of other lanthanides such as Eu$^{3+}$ \cite{Drexhage1970, CPS1978, Amos1997} and Er$^{3+}$ \cite{Snoeks1995, Kippenberg2006, Kroekenstoel2009}. However, from those experiments, the maximum emission rate enhancement of 2.4 folds was obtained through the coupling between Er$^{3+}$ ions and plasmonic cavity \cite{Kroekenstoel2009}. Here we demonstrate the maximum emission rate enhancement of 3.8 folds for the simple coupling between YAG:Ce$^{3+}$ nanocrystals with the gold NW}. After collecting four samples, we obtain an average of emission rate enhancements of 2.9 folds. This is still much larger than that of 1.7 folds measured for fluorescent dyed spheres, which also measured previously by Frimmer et al. \cite{Frimmer2011}. For explaining both observations of YAG:Ce$^{3+}$ nanocrystals and fluorescent dyed spheres, we model our single light sources in the proximity of the gold NW with a combination of the lifetime calculations from distribution of emitters inside a sphere based on the analytical model of an electric dipole \cite{Chew1988} and a toolbox for the simulation of metallic nanoparticles (MNP), using a boundary element method (BEM) approach \cite{Hohenester2012}.       

\section{Experiments and Results}

YAG:Ce$^{3+}$ (0.5 mol$\%$) nanocrystals were prepared by using a combustion synthesis technique \cite{Wuister2004}, which was optimized to yield nanoparticles of about $\sim$ 20 nm with a small degree of sintering. However, they usually formed micrometer-size clusters as shown in Fig. \ref{Figure1}a. The YAG:Ce$^{3+}$ clusters are highly friable and can be easily broken apart into myriad of nanocrystals by sonic waves (sonic bath) for half an hour. The transmission electron microscopy (TEM) image from pristine YAG:Ce$^{3+}$ nanocrystals is depicted in Fig. \ref{Figure1}b. Since the nanocrystals are not perfect spheres, we substituted the diameter in our model with the average of the diameter ($d_{lsor}$) of equivalent prolate spheroids \cite{Birowosuto2010}. The inset in Fig. \ref{Figure1}b exhibits the histogram of $d_{lsor}$ and the average of $d_{lsor}$ is 22 $\pm$ 10 nm obtained via Fast Fourier Transform (FFT) of five TEM images. Then, we prepared two polymer-coated samples, i.e. YAG:Ce$^{3+}$ nanocrystals reference and YAG:Ce$^{3+}$ nanocrystals-gold NW samples as shown in left and right sides of Fig. \ref{Figure1}c, respectively. For all samples, we prepared one suspension of 1 ml Polyvinyl Alcohol (PVA) 1$\%$ mixed with 0.1 ml diluted solution of YAG:Ce$^{3+}$ nanocrystals and one solution of 2 ml (Poly)methyl Methacrylate (PMMA) 5$\%$. As for the reference, the 0.1 ml suspension of YAG:Ce$^{3+}$ nanocrystals was spun on the cover slide at 6,000 rpm for 20 seconds resulting in $\sim$20-30 nm thickness layer \textcolor{red}{with a low areal density of about 0.6 $\mu$m$^{-2}$ \cite{Birowosuto2010}}. Then, on the top of the first layer, the 0.2 ml suspension of PMMA was spun at 1,000 rpm for 20 seconds resulting in 1 $\mu$m thickness layer. In the preparation for the YAG:Ce$^{3+}$ nanocrystals-gold NW samples, the suspension of PMMA was mixed with gold NW (Nanopartz, diameter $d_{nw}$ = 100 nm, length $l_{nw}$ = 2 $\mu$m) before the suspension was spun on the top of the PVA layer. The thickness of the resulting PVA layer was characterized by atomic force microscopy spanning a region containing a scratch made on the layer with a Teflon tweezers, yielding a homogeneous layer of $t$ = 21 $\pm$ 3 nm, see Fig. \ref{Figure1}d. The distribution range of distances between the Ce$^{3+}$ ions inside YAG nanocrystal and the gold NW, used in the analysis, is obtained from the diameter of the light source $d_{lsor}$. For the comparison, we also measured the samples with fluorescent dyed spheres \textcolor{red}{with the same particle density as YAG:Ce$^{3+}$ nanocrystals}. The sphere is a polystyrene bead with a diameter $d_{lsor}$ of 100 nm, infiltrated with approximately 1,000 randomly oriented dye molecules, a fluorescence peak at 560 nm, and a quantum efficiency close to 1 (Invitrogen Fluospheres F8800) \cite{Frimmer2011}. \textcolor{red}{In fact, we measured similar fluorescent dyed spheres with the same quantum efficiency \cite{Birowosuto2010} by positioning the spheres at precisely defined distances from the mirror to control the LDOS \cite{Cesa2009}}.    

Photoluminescence (PL), bleaching, and time-resolved emission measurements were performed at room temperature. The sample was optically excited using the continuous-wave or 10-MHz pulse diode laser at 450 nm with an excitation power of 1 $\mu$W \textcolor{red}{and a beam spot diameter of 2 $\mu$m}. For bleaching and time resolved emission, we used the pulse diode laser and filtered the emission with a Semrock Brightline at a maximum peak of 590 nm and 104-nm width. Fig. \ref{Figure2}a shows the PL spectrum of single YAG:Ce$^{3+}$ nanocrystals inside reference samples. The emission is dominated by Ce$^{3+}$ that consists of two main overlapping bands at 525 nm and 572 nm. They correspond to transitions from the lowest 5d level to $4f^{1}[^{2}F_{5/2}]$ and $4f^{1}[^{2}F_{7/2}]$ levels on Ce$^{3+}$ ions \cite{Birowosuto2009}. The anomalous broad band at longer wavelengths in the spectrum can be related with the defect of Ce$^{3+}$ emission \cite{Wong1984}. In Fig. \ref{Figure2}b, we did not observe bleaching for single YAG:Ce$^{3+}$ nanocrystals in thirty minutes. In fact, in the same time scale, we observed the loss of the half of the intensities for 100-nm diameter fluorescent dyed spheres.       

Then, we investigated YAG:Ce$^{3+}$ nanocrystals-gold NW samples. From the intensities, we located and selected four YAG:Ce$^{3+}$ nanocrystals, which are in the proximity of the gold NW. Microscope and PL images of single YAG:Ce$^{3+}$ nanocrystals are shown in Fig. \ref{Figure3}. From the microscope image of Fig. \ref{Figure3}a, we located single YAG:Ce$^{3+}$ nanocrystals. We observe that single nanocrystals close to the gold NW have two-time higher intensities in comparison with that alone, see PL images in Fig. \ref{Figure3}b and c. For the single nanocrystals close to the gold NW, we also observed the emission coupling of YAG:Ce$^{3+}$ nanocrystals to the propagating plasmonic mode of the gold NW. \cite{Akimov2007,Kolesov2009,Frimmer2011,AlexHuck2011}.      

Finally, we measured the emission decay curves of single YAG:Ce$^{3+}$ nanocrystals from the reference samples and the samples which the nanocrystals  coupled with the gold NW, see Fig. \ref{Figure4}a. The decay curve of single YAG:Ce$^{3+}$ nanocrystals in the reference samples exhibit a single exponential with a fit through the curve yields a lifetime ($t_{hom}$) of 60 ns. This lifetime was already attributed previously for 5d-4f Ce$^{3+}$ emission\cite{Wuister2004} and it is 10-ns shorter than the 70-ns decay time reported in YAG:Ce$^{3+}$ bulk crystals \cite{Wong1984}. For the single nanocrystals close to the gold NW, the decay curve is nonexponential. The curve was fitted with two exponential yielding a fast decay component 3.6 ns (10$\%$) and a slow component of 17.1 ns (90$\%$). The average emission lifetime ($t_{goldNW}$) is then 15.8 ns. 
\textcolor{red}{This emission lifetime is much faster than that in YAG:Ce$^{3+}$ nanocrystals without the presence of gold NW. We expect that this lifetime reduction is spatially extended along the NW radius as it was theoretically predicted that the emission rate enhancement occurs mostly due to the coupling of the emitter to guided plasmonic mode of NW \cite{DEChang2006}. The intensity of the decay curve from the single YAG:Ce$^{3+}$ nanocrystals close to the gold NW is lower than that in the reference. The low detection efficiency for the YAG:Ce$^{3+}$ nanocrystals close to gold NW seems to be the problem. The photons from the emission of the single YAG:Ce$^{3+}$ nanocrystals are strongly scattered by the gold NW located above and those photons are unlikely to be in the direction of the light collection.} 

From our experiments, we derived the emission rate enhancement factor ($t_{hom}/t_{goldNW}$) of 3.8. Four enhancement data and those for 100-nm diameter fluorescent dyed spheres are compiled as box charts in Fig. \ref{Figure4}b. From the charts, the emission rate enhancements for YAG:Ce$^{3+}$ nanocrystals and fluorescent dyed spheres, which almost have the same emission wavelength, are 2.9 $\pm$ 0.7 and 1.7 $\pm$ 0.3, respectively. The enhancements for fluorescent dyed spheres are the same with the earlier observation \cite{Frimmer2011}. The larger enhancements in YAG:Ce$^{3+}$ nanocrystals to those in fluorescent dyed spheres can be strongly correlated either with the refractive index or the diameter of the light sources. In the following discussions, we will investigate both factors with the calculations using the analytical model of spherical particle scattering and MNPBEM. 

\section{Discussions}

For the investigation on the effects of the refractive index and diameter of the light source to the emission rate enhancements, we propose a single sphere plasmonic nanowire model as a combination of the analytical model of spontaneous emission of an electric dipole located inside a spherical particle \textcolor{red}{as defined previously by Chew} \cite{Chew1988} and BEM simulation of the spontaneous emission near a plasmonic NW using the solution of the full Maxwell equations \cite{Hohenester2012}.
The scheme of the calculation and simulation is illustrated in the inset of Fig. \ref{Figure5}a. First, we modeled a YAG:Ce$^{3+}$ nanocrystal or a fluorescent dyed sphere as a spherical particle with a refractive index $n_{lsor}$ and a dielectric constant $\epsilon_{lsor}$ uniformly filled with Ce$^{3+}$ ion or dye molecules inside a host medium of polymer, with a refractive index $n_{poly}$ and a dielectric constant $\epsilon_{poly}$ \cite{Chew1988}. The emission rate of single Ce$^{3+}$ ion or single dye at specific location with radial coordinates $r'$ inside the single \textcolor{red}{nanocrystal or fluorescent dyed sphere, respectively,} is given by \cite{Chew1988}:
\begin{equation}
\begin{aligned}
R^{\perp}/R^{\perp}_{0} &= \frac{3\epsilon_{lsor}n_{lsor}}{2\rho_{lsor}^{2}}\left[\epsilon_{poly}\right]^{1/2} \times \\
&\Sigma^{\infty}_{n = 1} n(n + 1)(2n + 1) \frac{j_{n}^{2}(y_{lsor})}{y_{lsor}^{2}|D_{n}|^{2}},
\end{aligned}
\label{eq1}
\end{equation}
for radial oscillations with the respect of the center of the \textcolor{red}{nanocrystal or fluorescent dyed sphere} and
\begin{equation}
\begin{aligned}
R^{\parallel}/R^{\parallel}_{0} &= \frac{3\epsilon_{lsor}n_{lsor}}{4\rho_{lsor}^{2}}\left[\epsilon_{poly}\right]^{1/2} \times \\
&\Sigma^{\infty}_{n = 1} (2n + 1) \times \\ 
&\left(\left|\frac{[y_{lsor}j_{n}(y_{lsor})]'}{y_{lsor}D_{n}}\right|^{2} +
\frac{1}{\epsilon_{lsor}\epsilon_{poly}}\frac{j_{n}^{2}(y_{lsor})}{|D_{n}'|^{2}}\right),
\end{aligned}
\label{eq2}
\end{equation}
for tangential oscillations. The terms with the factor $D_{n}$ are electric multipole (TM) terms and those with the factor $D_{n}'$ are magnetic multiple (TE) terms. Some related parameters are
\begin{equation}
\begin{aligned}
n_{lsor} &= \sqrt{\epsilon_{lsor}}, y_{lsor} = k_{lsor}r', \\
\rho_{lsor,poly} &= k_{lsor,poly}d_{lsor}, \\
D_{n} &= \epsilon_{lsor},j_{n}(\rho_{lsor})[\rho_{poly}h_{n}^{lsor}(\rho_{poly})]' \\
        &-\epsilon_{poly}h_{n}^{lsor}(\rho_{poly})[\rho_{lsor}j_{n}(\rho_{lsor})]', \\
D_{n}' &= D_{n} \textrm{~with~} \epsilon_{lsor,poly} \rightarrow 1,
\end{aligned}
\label{eq3}
\end{equation}
The denominators $D_{n}$ and $D_{n}$' are the same as those of the elastic (Mie) scattering coefficients \cite{Chew1988}. The formulation above is described for mapping the arbitrary distribution of the emission rate for single Ce$^{3+}$ ions or single dyes inside the \textcolor{red}{nanocrystal or fluorescent dyed sphere, respectively}. However, there is a limitation in the calculation for the emission rate of the single Ce$^{3+}$ ions or single dyes in the proximity close to the surface of the \textcolor{red}{nanocrystal or fluorescent dyed sphere, respectively,} and the plasmonic nanowire. Therefore, we should define the average emission rates for that reason and give analytic and numerical results for the special case of a uniform distribution of Ce$^{3+}$ ions or dyes inside the \textcolor{red}{nanocrystal or fluorescent dyed sphere, respectively}. In details, we introduce the normalized average emission rate for either radial \textcolor{red}{$\left<R^{\perp}/R_{0}^{\perp}\right>$} or tangential oscillations \textcolor{red}{$\left<R^{\parallel}/R_{0}^{\parallel}\right>$} to be the total normalized radiated power for one of the polarizations divided by the total number of excited Ce$^{3+}$ ions or dyes \cite{Chew1988}.
\begin{equation}
\left<R^{\perp,\parallel}/R_{0}^{\perp,\parallel}\right> = \int{\left(R^{\perp,\parallel}/R_{0}^{\perp,\parallel}\right)n(\textbf{r'})d^{3}r'}/ \int{n(\textbf{r'})d^{3}r'},
\label{eq4}
\end{equation}    
where $R^{\perp}/R_{0}^{\perp}$ or $R^{\parallel}/R_{0}^{\parallel}$ is given by Eq. \ref{eq1} or Eq. \ref{eq2}, respectively. $n(\textbf{r})$ is the density of excited \textcolor{red}{Ce$^{3+}$ ions or dyes inside the nanocrystal or fluorescent dyed sphere, respectively}. For a uniform distribution, $n(\textbf{r})$ is a constant while the averages reduce to 
\begin{equation}
\begin{aligned}
\left<R^{\perp}/R_{0}^{\perp}\right> &= (3/4{\pi}d_{lsor}^{3})\int(R^{\perp}/R_{0}^{\perp})d^{3}r' \\
&= 2H \Sigma_{n=1}^{\infty} n(n + 1)L_{n}|1D_{n}|^{2},
\end{aligned}
\label{eq5}
\end{equation} 
\begin{equation}
\begin{aligned}
\left<R^{\parallel}/R_{0}^{\parallel}\right> &= (3/4{\pi}d_{lsor}^{3})\int(R^{\parallel}/R_{0}^{\parallel})d^{3}r' \\
&= H \Sigma_{n=1}^{\infty} \left(\frac{M_{n}}{|D_{n}|^{2}}+\frac{(2n + 1)PK_{n}}{|D_{n}'|^{2}}\right),
\end{aligned}
\label{eq6}
\end{equation}
where
\begin{equation}
\begin{aligned}
P &= 1/(\epsilon_{lsor}\epsilon_{poly}), \\
H &= (9\epsilon_{lsor}/4\rho_{lsor}^{5})(\epsilon_{lsor}\epsilon_{poly})^{1/2}, \\
K_{n} &= \int_{0}^{\rho_{lsor}}\rho^{2}j_{n}^{2}(\rho)d\rho, \\
&= (\rho_{lsor}^{3}/2)\left[j_{n}^{2}(\rho_{lsor})-j_{n+1}(\rho_{lsor})j_{n-1}(\rho_{lsor})\right], \\ 
L_{n} &= (2n + 1) \int_{0}^{\rho_{1}}j_{n}^{2}(\rho)d\rho, \\
M_{n} &= (2n + 1) \int_{0}^{\rho_{1}}\{[\rho{j_{n}(\rho)}]'\}^{2}d{\rho}.
\end{aligned}
\label{eq7}
\end{equation}
The integrals $L_{n}$ dan $M_{n}$ can be evaluated by the process of recursion using both spherical Bessel functions and the sine integral \cite{Chew1988}. 
\textcolor{red}{Since we have an interest only for the total rate average over the polarizations, we calculate $R/R_{0} = (R^{\perp}/R^{\perp}_{0})/3+2(R^{\parallel}/R^{\parallel}_{0})/3$ and $\left<R/R_{0}\right> = \left<R^{\perp}/R^{\perp}_{0}\right>/3 +2\left<R^{\parallel}/R^{\parallel}_{0}\right>/3$ for Eqs. \ref{eq1} and \ref{eq2} and Eqs. \ref{eq5} and \ref{eq6}, respectively.}

Then, we employed MNPBEM simulation toolbox \cite{Hohenester2012} using the emission rates from the analytical model, see the inset in Fig. \ref{Figure5}a. We used this approach instead of another analytical NW scattering model \cite{Birowosuto2014a} since the metallic nanowire plasmons are much more complicated and they depend strongly on the particle geometry and the interparticle coupling with the light source geometry. Thus, the MNPBEM simulation toolbox in our model computed for a given external perturbation of the induced electromagnetic fields created by a nearby emitter. This was achieved by solving full Maxwell equations and using the boundary condition at the metallic nanowire boundaries. \textcolor{red}{For the calculation of the emission rate, we limit the distance between the Ce$^{3+}$ ions or dyes and the gold NW as close as 0.5 nm. This distance can be attributed to the surface wall of the single nanocrystal or the single fluorescent dyed sphere.}

For the solutions with full Maxwell equations, we need both scalar $\phi$ and vector $\mathbf{A}$ potentials while both potentials are related through the Lorentz gauge condition $\nabla{\cdot}\mathbf{A}=ik\epsilon\phi$ \cite{FJAbajo2002}. A convenient solution scheme is given by the Green function of the wave equation.
\begin{equation}
\begin{aligned}
(\nabla^{2}+k_{i}^{2})G_{i}(\mathbf{r},\mathbf{r'}) &= -4\pi\delta(\mathbf{r}-\mathbf{r'}), \\
\textrm{~with~} G_{i}(\mathbf{r},\mathbf{r'}) &= \frac{exp({ik_{i}|r-r'|})}{|r-r'|} \textrm{~and~} i \rightarrow lsor, poly,
\end{aligned}
\label{eq8}
\end{equation}
where $k_{i} = \sqrt{\epsilon_{i}}k$ is the wavenumber in the medium outside the nanowire, $k = {\omega}/c$ is the wavenumber in vacuum, and $c$ is the speed of light. In our model, the medium outside the nanowire consists of the spherical particle and the polymer medium and we therefore may replace index $i$ with ${lsor}$ and ${poly}$, respectively. As consequences, we can write down the solutions in the forms:
\begin{equation}
\begin{aligned}
\phi(\mathbf{r}) &= \phi_{ext}(\mathbf{r}) + \oint_{V_{lsor,poly}}G_{lsor,poly}(\mathbf{r},\mathbf{s})\sigma_{lsor,poly}(\mathbf{s})da,
\end{aligned}
\label{eq9}
\end{equation}
and
\begin{equation}
\begin{aligned}
\mathbf{A}(\mathbf{r}) &= \mathbf{A}_{ext}(\mathbf{r}) + \oint_{V_{lsor,poly}}G_{lsor,poly}(\mathbf{r},\mathbf{s})\mathbf{h}_{lsor,poly}(\mathbf{s})da.
\end{aligned}
\label{eq10}
\end{equation}
Both solutions fulfill the Helmholtz equations everywhere except at the particle boundaries. $\sigma_{lsor}$, $\sigma_{poly}$, $h_{lsor}$, and $h_{poly}$ are surface charges and current distributions inside the spherical particle and the polymer medium, respectively. $\phi_{ext}$ and $\mathbf{A}_{ext}$ are the scalar and vector potentials characterizing the external perturbation. In this MNPBEM approach, the integrals derived from Eqs. \ref{eq9} and \ref{eq10} are approximated by sums over boundary elements. Working on the boundary conditions of Maxwell equations, we derived a set of equations for $\sigma_{lsor,poly}$ and $h_{lsor,poly}$ \cite{FJAbajo2002}, which could be solved through matrix inversions and multiplications. After solving $\sigma_{lsor,poly}$ and $h_{lsor,poly}$, we computed the potential everywhere else through Eqs. \ref{eq9} and \ref{eq10} as well as the electromagnetic fields. Those fields are related to the potentials through the usual relations $\mathbf{E} = ik\mathbf{A} - \nabla{\phi}$ and $\mathbf{H} = \nabla \times \mathbf{A}$.  

In summary, using the analytical model, we first obtained the emission rates from Eqs. \ref{eq1},\ref{eq2},\ref{eq5},and \ref{eq6} with $d_{lsor}$ and $n_{lsor}$ as independent variables. After that, we employed MNPBEM simulations with parameters of the single gold NW, i.e. the diameter $d_{nw}$ = 100 nm, the length $l_{nw}$ = 2 ~$\mu$m, and $\epsilon_{nw}$ = $-8.29 + 1.97i$. Fig. \ref{Figure5}a shows the distribution of the emission rate enhancements inside the single spheres in the proximity of the gold NW as models for the YAG:Ce$^{3+}$ nanocrystals ($n_{lsor}$ = 1.83) and fluorescent dyed spheres ($n_{lsor}$ = 1.59) in the experiments. For the distribution, the emission rates from Eqs. \ref{eq1} and \ref{eq2} were used as inputs for MNPBEM simulations. However, the rates calculated in Fig. \ref{Figure5}a are limited only for the rates at a distance of 0.5 nm from the surfaces of the spheres. For YAG:Ce$^{3+}$ nanocrystals with a diameter of 22 nm, the relative variance of the emission rate enhancements $s^{2}/\bar{x}^{2}$ is 0.009. This variance is twenty times smaller than that of single fluorescent dyed sphere with a diameter of 100 nm of 0.167. The maximum and the minimum enhancements of six and one folds, respectively, is only achieved for dyes inside single fluorescent dyed sphere. 

After the emission rate distribution is discussed, we focus on the average emission rates for the ensemble emitters as those were actually measured in our experiments. For the analysis of the average emission enhancements from the ensemble Ce$^{3+}$ doped ions and dyes inside YAG nanocrystals or fluorescent dyed spheres, respectively, the average emission rates from Eqs. \ref{eq5} and \ref{eq6} were used as inputs for MNPBEM simulation. Then, we also vary $d_{lsor}$ and $n_{lsor}$ as attempts to obtain information about the effects of the light source sizes and refractive indexes to the emission enhancements. The results are summarized in Figs. \ref{Figure5}b and \ref{Figure5}c. First, we discuss the emission rates for the light source solely inside the polymer medium $n_{poly} = 1.55$. For the same sizes of the light sources shown in Fig. \ref{Figure5}b, the emission rates get more inhibited when the refractive indexes get larger \cite{Chew1988}. However, the emission rates are more complicated for different light source sizes. In one case, the emission rates for small size light source tend to get more enhanced for the refractive indexes of the light sources smaller than the polymer medium ($n_{lsor} < n_{poly}$). In another case, those tend to get more inhibited for $n_{lsor} > n_{poly}$, see Fig. \ref{Figure5}c. In addition, as the refractive index contrast between the light source with the polymer gets larger and the size becomes comparable with the wavelength, we observe the resonance of the single sphere as shown in the inset of Fig. \ref{Figure5}b. 

Finally, we analyze the emission rates for the light source near the gold NW. As shown in Fig. \ref{Figure5}b, the resonance now completely disappears as the enhancement of the emission rates due to the metallic nanowire is strong. However, although the plasmonic effect is large, the trends for the different sizes and the different refractive indexes of the light sources are still the same as those in the single sphere only, see Figs. \ref{Figure5}b and \ref{Figure5}c. If the refractive indexes $n_{lsor}$ are 1.83 (YAG:Ce$^{3+}$ nanocrystals) and 1.59 (fluorescent dyed spheres), the size of YAG:Ce$^{3+}$ nanocrystals with large $n_{lsor}$ are expected to be half of the fluorescent spheres if both have the same emission rates. In a relationship with our experiments, the statistic values in Fig. \ref{Figure4}b indicated as solid circles in Figs. \ref{Figure5}b and \ref{Figure5}c are in agreement with the calculations. Therefore, we emphasize that the ions in 22-nm-diameter YAG:Ce$^{3+}$ nanocrystals still have large emission rate enhancements in comparison to the dyes in 100-nm-diameter fluorescent spheres although the large refractive index of YAG:Ce$^{3+}$ nanocrystals gives small contribution to the inhibition of the emission rates.              

\section{Conclusion}

Here we reported a nano-assembled system comprising single YAG:Ce$^{3+}$ nanocrystals and single gold NWs. The maximum and the average emission rate enhancements of 3.8 and  2.9 $\pm$ 0.7 folds were reported respectively. We also compared the emission rate enhancements of the fluorescent dyed spheres in the same NW system with the same emission wavelength. The emission rate enhancement for YAG:Ce$^{3+}$ nanocrystals is almost twice of that for fluorescent dyed spheres. Using the analytical scattering model and BEM method, we calculated the average emission of ensemble Ce$^{3+}$ ions and dye molecules inside single nanocrystals and fluorescent spheres, respectively. We found that the 22-nm-small-size YAG:Ce$^{3+}$ nanocrystals are more efficient for the coupling with the gold NW in comparison with fluorescent dyed spheres. Additionally, YAG:Ce$^{3+}$ nanocrystals are photostable, which make these NW-emitter systems attractive for the applications of nanoscale light sources while those of solid state lighting and display devices are tantalizing.  

\section{Acknowledgments}
M. D. B. would like to acknowledge University of Twente for the early helps on the project and Isnaeni would like to acknowledge the Research of Excellence funding of Indonesian Institute of Sciences.

\newpage

\begin{figure}[htbp]
\caption{(a) Dark-field microscope image of Cerium doped Yttrium Aluminum Garnet (YAG:Ce$^{3+}$) nanocrystals microclusters. (b) Transmission emission microscope image of the nanocrystals after the cluster-separation process and histogram of size distribution of nanocrystals (inset). (c) Reference (left) and gold-NW-embedded (right) samples of nanocrystals inside Polyvinil Alcohol (PVA) dan (Poly)methyl Methacrylate (PMMA). (d) Cross section curves from the indicated color lines in the atomic force microscope image (inset).}
\label{Figure1}
\end{figure}

\begin{figure}[htbp]
\caption{(a) Measured photoluminescence (PL) spectrum and (b) bleaching properties of single YAG:Ce$^{3+}$ nanocrystals inside reference samples. The dotted lines in (a) and (b) indicates two bands from the transitions from the lowest 5d level to $4f^{1}[^{2}F_{5/2}]$ and $4f^{1}[^{2}F_{7/2}]$ levels and the bleaching curve of fluorescent dyed spheres, respectively.}
\label{Figure2}
\end{figure}

\begin{figure}[htbp]
\caption{(a) Microscope image of the YAG:Ce$^{3+}$ nanocrystals with gold NW sample. (b, c) PL images of the single nanocrystals close in the proximity of the single gold NW (b) and that of single nanocrystals alone (c). The circled area in (a) with the letters correspond to the same nanocrystals in PL images.}
\label{Figure3}
\end{figure}

\begin{figure}[htbp]
\caption{(a) PL decay curves of the YAG:Ce$^{3+}$ nanocrystals in the reference sample (black filled circles) and close to the gold NW (red empty circles). The curves were normalized to that of black filled circles for clarity. (b) Comparison emission rate enhancements for an emitter close to 2-$\mu$m-length gold NW between Cerium doped YAG nanocrystals and 100-nm diameter fluorescent dyed spheres. Emission in all experiments was filtered at 590 nm.}
\label{Figure4}
\end{figure}

\begin{figure}[htbp]
\caption{Calculations of the emission rate of the finite-size light sources in the proximity of gold NW. (a) Distribution of the emission rate enhancement inside single light sources by considering the size and the refractive index. Average emission rate enhancement for different (b) diameter $d_{lsor}$ and (c) refractive index of the single light sources $n_{lsor}$. Solid circles indicate the experimental values in Fig. \ref{Figure4}b. The inset in (b) shows the observation of the resonances for the large-diameter light sources. In this calculations, the single light sources were modeled as single spheres. The inset in (a) shows the calculated average emission rate of the dipole inside a sphere and its combination with MNPBEM simulation for the emission rate close to a gold NW.}
\label{Figure5}
\end{figure}


\begin{thebibliography}{33}%
\makeatletter
\providecommand \@ifxundefined [1]{%
 \@ifx{#1\undefined}
}%
\providecommand \@ifnum [1]{%
 \ifnum #1\expandafter \@firstoftwo
 \else \expandafter \@secondoftwo
 \fi
}%
\providecommand \@ifx [1]{%
 \ifx #1\expandafter \@firstoftwo
 \else \expandafter \@secondoftwo
 \fi
}%
\providecommand \natexlab [1]{#1}%
\providecommand \enquote  [1]{``#1''}%
\providecommand \bibnamefont  [1]{#1}%
\providecommand \bibfnamefont [1]{#1}%
\providecommand \citenamefont [1]{#1}%
\providecommand \href@noop [0]{\@secondoftwo}%
\providecommand \href [0]{\begingroup \@sanitize@url \@href}%
\providecommand \@href[1]{\@@startlink{#1}\@@href}%
\providecommand \@@href[1]{\endgroup#1\@@endlink}%
\providecommand \@sanitize@url [0]{\catcode `\\12\catcode `\$12\catcode
  `\&12\catcode `\#12\catcode `\^12\catcode `\_12\catcode `\%12\relax}%
\providecommand \@@startlink[1]{}%
\providecommand \@@endlink[0]{}%
\providecommand \url  [0]{\begingroup\@sanitize@url \@url }%
\providecommand \@url [1]{\endgroup\@href {#1}{\urlprefix }}%
\providecommand \urlprefix  [0]{URL }%
\providecommand \Eprint [0]{\href }%
\providecommand \doibase [0]{http://dx.doi.org/}%
\providecommand \selectlanguage [0]{\@gobble}%
\providecommand \bibinfo  [0]{\@secondoftwo}%
\providecommand \bibfield  [0]{\@secondoftwo}%
\providecommand \translation [1]{[#1]}%
\providecommand \BibitemOpen [0]{}%
\providecommand \bibitemStop [0]{}%
\providecommand \bibitemNoStop [0]{.\EOS\space}%
\providecommand \EOS [0]{\spacefactor3000\relax}%
\providecommand \BibitemShut  [1]{\csname bibitem#1\endcsname}%
\let\auto@bib@innerbib\@empty
\bibitem [{\citenamefont {Nirmal}\ \emph {et~al.}(1996)\citenamefont {Nirmal},
  \citenamefont {Dabbousi}, \citenamefont {Bawendi}, \citenamefont {Macklin},
  \citenamefont {Trautman}, \citenamefont {Harris},\ and\ \citenamefont
  {Brus}}]{Nirmal1996}%
  \BibitemOpen
  \bibfield  {author} {\bibinfo {author} {\bibfnamefont {M.}~\bibnamefont
  {Nirmal}}, \bibinfo {author} {\bibfnamefont {B.}~\bibnamefont {Dabbousi}},
  \bibinfo {author} {\bibfnamefont {M.}~\bibnamefont {Bawendi}}, \bibinfo
  {author} {\bibfnamefont {J.}~\bibnamefont {Macklin}}, \bibinfo {author}
  {\bibfnamefont {J.}~\bibnamefont {Trautman}}, \bibinfo {author}
  {\bibfnamefont {T.}~\bibnamefont {Harris}}, \ and\ \bibinfo {author}
  {\bibfnamefont {L.}~\bibnamefont {Brus}},\ }\href@noop {} {\bibfield
  {journal} {\bibinfo  {journal} {Nature}\ }\textbf {\bibinfo {volume} {383}},\
  \bibinfo {pages} {802} (\bibinfo {year} {1996})}\BibitemShut {NoStop}%
\bibitem [{\citenamefont {Fleury}\ \emph {et~al.}(1998)\citenamefont {Fleury},
  \citenamefont {Sick}, \citenamefont {Zumofen}, \citenamefont {Hecht},\ and\
  \citenamefont {Wild}}]{Fleur1998}%
  \BibitemOpen
  \bibfield  {author} {\bibinfo {author} {\bibfnamefont {L.}~\bibnamefont
  {Fleury}}, \bibinfo {author} {\bibfnamefont {B.}~\bibnamefont {Sick}},
  \bibinfo {author} {\bibfnamefont {G.}~\bibnamefont {Zumofen}}, \bibinfo
  {author} {\bibfnamefont {B.}~\bibnamefont {Hecht}}, \ and\ \bibinfo {author}
  {\bibfnamefont {W.~P.}\ \bibnamefont {Wild}},\ }\href@noop {} {\bibfield
  {journal} {\bibinfo  {journal} {Molecular Physics}\ }\textbf {\bibinfo
  {volume} {95}},\ \bibinfo {pages} {1333} (\bibinfo {year}
  {1998})}\BibitemShut {NoStop}%
\bibitem [{\citenamefont {Birowosuto}\ and\ \citenamefont
  {Dorenbos}(2009)}]{Birowosuto2009}%
  \BibitemOpen
  \bibfield  {author} {\bibinfo {author} {\bibfnamefont {M.~D.}\ \bibnamefont
  {Birowosuto}}\ and\ \bibinfo {author} {\bibfnamefont {P.}~\bibnamefont
  {Dorenbos}},\ }\href@noop {} {\bibfield  {journal} {\bibinfo  {journal}
  {physica status solidi (a)}\ }\textbf {\bibinfo {volume} {206}},\ \bibinfo
  {pages} {9} (\bibinfo {year} {2009})}\BibitemShut {NoStop}%
\bibitem [{\citenamefont {Wong}, \citenamefont {Rotman},\ and\ \citenamefont
  {Warde}(1984)}]{Wong1984}%
  \BibitemOpen
  \bibfield  {author} {\bibinfo {author} {\bibfnamefont {C.~M.}\ \bibnamefont
  {Wong}}, \bibinfo {author} {\bibfnamefont {S.~R.}\ \bibnamefont {Rotman}}, \
  and\ \bibinfo {author} {\bibfnamefont {C.}~\bibnamefont {Warde}},\
  }\href@noop {} {\bibfield  {journal} {\bibinfo  {journal} {Appl. Phys.
  Lett.}\ }\textbf {\bibinfo {volume} {44}} (\bibinfo {year}
  {1984})}\BibitemShut {NoStop}%
\bibitem [{\citenamefont {Mihóková}\ \emph {et~al.}(2007)\citenamefont
  {Mihóková}, \citenamefont {Nikl}, \citenamefont {Mareš}, \citenamefont
  {Beitlerová}, \citenamefont {Vedda}, \citenamefont {Nejezchleb},
  \citenamefont {Blažek},\ and\ \citenamefont {D’Ambrosio}}]{Nikl2007}%
  \BibitemOpen
  \bibfield  {author} {\bibinfo {author} {\bibfnamefont {E.}~\bibnamefont
  {Mihóková}}, \bibinfo {author} {\bibfnamefont {M.}~\bibnamefont {Nikl}},
  \bibinfo {author} {\bibfnamefont {J.}~\bibnamefont {Mareš}}, \bibinfo
  {author} {\bibfnamefont {A.}~\bibnamefont {Beitlerová}}, \bibinfo {author}
  {\bibfnamefont {A.}~\bibnamefont {Vedda}}, \bibinfo {author} {\bibfnamefont
  {K.}~\bibnamefont {Nejezchleb}}, \bibinfo {author} {\bibfnamefont
  {K.}~\bibnamefont {Blažek}}, \ and\ \bibinfo {author} {\bibfnamefont
  {C.}~\bibnamefont {D’Ambrosio}},\ }\href@noop {} {\bibfield  {journal}
  {\bibinfo  {journal} {Journal of Luminescence}\ }\textbf {\bibinfo {volume}
  {126}},\ \bibinfo {pages} {77 } (\bibinfo {year} {2007})}\BibitemShut
  {NoStop}%
\bibitem [{\citenamefont {Pan}, \citenamefont {Wu},\ and\ \citenamefont
  {Su}(2004)}]{Pan2004}%
  \BibitemOpen
  \bibfield  {author} {\bibinfo {author} {\bibfnamefont {Y.}~\bibnamefont
  {Pan}}, \bibinfo {author} {\bibfnamefont {M.}~\bibnamefont {Wu}}, \ and\
  \bibinfo {author} {\bibfnamefont {Q.}~\bibnamefont {Su}},\ }\href@noop {}
  {\bibfield  {journal} {\bibinfo  {journal} {Journal of Physics and Chemistry
  of Solids}\ }\textbf {\bibinfo {volume} {65}},\ \bibinfo {pages} {845 }
  (\bibinfo {year} {2004})}\BibitemShut {NoStop}%
\bibitem [{\citenamefont {Wuister}, \citenamefont {de~Mello~Donega},\ and\
  \citenamefont {Meijerink}(2004)}]{Wuister2004}%
  \BibitemOpen
  \bibfield  {author} {\bibinfo {author} {\bibfnamefont {S.~F.}\ \bibnamefont
  {Wuister}}, \bibinfo {author} {\bibfnamefont {C.}~\bibnamefont
  {de~Mello~Donega}}, \ and\ \bibinfo {author} {\bibfnamefont {A.}~\bibnamefont
  {Meijerink}},\ }\href@noop {} {\bibfield  {journal} {\bibinfo  {journal}
  {Phys. Chem. Chem. Phys.}\ }\textbf {\bibinfo {volume} {6}},\ \bibinfo
  {pages} {1633} (\bibinfo {year} {2004})}\BibitemShut {NoStop}%
\bibitem [{\citenamefont {Sprik}, \citenamefont {VanTiggelen},\ and\
  \citenamefont {Lagendijk}(1996)}]{Sprik1996}%
  \BibitemOpen
  \bibfield  {author} {\bibinfo {author} {\bibfnamefont {R.}~\bibnamefont
  {Sprik}}, \bibinfo {author} {\bibfnamefont {B.~A.}\ \bibnamefont
  {VanTiggelen}}, \ and\ \bibinfo {author} {\bibfnamefont {A.}~\bibnamefont
  {Lagendijk}},\ }\href@noop {} {\bibfield  {journal} {\bibinfo  {journal}
  {Europhys. Lett.}\ }\textbf {\bibinfo {volume} {35}},\ \bibinfo {pages} {265}
  (\bibinfo {year} {1996})}\BibitemShut {NoStop}%
\bibitem [{\citenamefont {Drexhage}(1970)}]{Drexhage1970}%
  \BibitemOpen
  \bibfield  {author} {\bibinfo {author} {\bibfnamefont {K.}~\bibnamefont
  {Drexhage}},\ }\href@noop {} {\bibfield  {journal} {\bibinfo  {journal} {J.
  Lumin.}\ }\textbf {\bibinfo {volume} {1 - 2}},\ \bibinfo {pages} {693 }
  (\bibinfo {year} {1970})}\BibitemShut {NoStop}%
\bibitem [{\citenamefont {Birowosuto}\ \emph
  {et~al.}(2014{\natexlab{a}})\citenamefont {Birowosuto}, \citenamefont
  {Zhang}, \citenamefont {Yokoo}, \citenamefont {Takiguchi},\ and\
  \citenamefont {Notomi}}]{Birowosuto2014a}%
  \BibitemOpen
  \bibfield  {author} {\bibinfo {author} {\bibfnamefont {M.~D.}\ \bibnamefont
  {Birowosuto}}, \bibinfo {author} {\bibfnamefont {G.}~\bibnamefont {Zhang}},
  \bibinfo {author} {\bibfnamefont {A.}~\bibnamefont {Yokoo}}, \bibinfo
  {author} {\bibfnamefont {M.}~\bibnamefont {Takiguchi}}, \ and\ \bibinfo
  {author} {\bibfnamefont {M.}~\bibnamefont {Notomi}},\ }\href@noop {}
  {\bibfield  {journal} {\bibinfo  {journal} {Opt. Express}\ }\textbf {\bibinfo
  {volume} {22}},\ \bibinfo {pages} {11713} (\bibinfo {year}
  {2014}{\natexlab{a}})}\BibitemShut {NoStop}%
\bibitem [{\citenamefont {Lodahl}\ \emph {et~al.}(2004)\citenamefont {Lodahl},
  \citenamefont {{Floris Van Driel}}, \citenamefont {Nikolaev}, \citenamefont
  {Irman}, \citenamefont {Overgaag}, \citenamefont {Vanmaekelbergh},\ and\
  \citenamefont {Vos}}]{Lodahl2004}%
  \BibitemOpen
  \bibfield  {author} {\bibinfo {author} {\bibfnamefont {P.}~\bibnamefont
  {Lodahl}}, \bibinfo {author} {\bibfnamefont {A.}~\bibnamefont {{Floris Van
  Driel}}}, \bibinfo {author} {\bibfnamefont {I.~S.}\ \bibnamefont {Nikolaev}},
  \bibinfo {author} {\bibfnamefont {A.}~\bibnamefont {Irman}}, \bibinfo
  {author} {\bibfnamefont {K.}~\bibnamefont {Overgaag}}, \bibinfo {author}
  {\bibfnamefont {D.}~\bibnamefont {Vanmaekelbergh}}, \ and\ \bibinfo {author}
  {\bibfnamefont {W.~L.}\ \bibnamefont {Vos}},\ }\href@noop {} {\bibfield
  {journal} {\bibinfo  {journal} {Nature}\ }\textbf {\bibinfo {volume} {430}},\
  \bibinfo {pages} {654} (\bibinfo {year} {2004})}\BibitemShut {NoStop}%
\bibitem [{\citenamefont {Birowosuto}\ \emph
  {et~al.}(2012{\natexlab{a}})\citenamefont {Birowosuto}, \citenamefont
  {Sumikura}, \citenamefont {Matsuo}, \citenamefont {Taniyama}, \citenamefont
  {van Veldhoven}, \citenamefont {Noetzel},\ and\ \citenamefont
  {Notomi}}]{Birowosuto2012}%
  \BibitemOpen
  \bibfield  {author} {\bibinfo {author} {\bibfnamefont {M.~D.}\ \bibnamefont
  {Birowosuto}}, \bibinfo {author} {\bibfnamefont {H.}~\bibnamefont
  {Sumikura}}, \bibinfo {author} {\bibfnamefont {S.}~\bibnamefont {Matsuo}},
  \bibinfo {author} {\bibfnamefont {H.}~\bibnamefont {Taniyama}}, \bibinfo
  {author} {\bibfnamefont {P.~J.}\ \bibnamefont {van Veldhoven}}, \bibinfo
  {author} {\bibfnamefont {R.}~\bibnamefont {Noetzel}}, \ and\ \bibinfo
  {author} {\bibfnamefont {M.}~\bibnamefont {Notomi}},\ }\href@noop {}
  {\bibfield  {journal} {\bibinfo  {journal} {{Sci. Rep.}}\ }\textbf {\bibinfo
  {volume} {{2}}},\ \bibinfo {pages} {{321}} (\bibinfo {year}
  {{2012}}{\natexlab{a}})}\BibitemShut {NoStop}%
\bibitem [{\citenamefont {Birowosuto}\ \emph
  {et~al.}(2012{\natexlab{b}})\citenamefont {Birowosuto}, \citenamefont
  {Yokoo}, \citenamefont {Taniyama}, \citenamefont {Kuramochi}, \citenamefont
  {Takiguchi},\ and\ \citenamefont {Notomi}}]{Birowosuto2012b}%
  \BibitemOpen
  \bibfield  {author} {\bibinfo {author} {\bibfnamefont {M.~D.}\ \bibnamefont
  {Birowosuto}}, \bibinfo {author} {\bibfnamefont {A.}~\bibnamefont {Yokoo}},
  \bibinfo {author} {\bibfnamefont {H.}~\bibnamefont {Taniyama}}, \bibinfo
  {author} {\bibfnamefont {E.}~\bibnamefont {Kuramochi}}, \bibinfo {author}
  {\bibfnamefont {M.}~\bibnamefont {Takiguchi}}, \ and\ \bibinfo {author}
  {\bibfnamefont {M.}~\bibnamefont {Notomi}},\ }\href@noop {} {\bibfield
  {journal} {\bibinfo  {journal} {J. of Appl. Phys.}\ }\textbf {\bibinfo
  {volume} {112}},\ \bibinfo {eid} {113106} (\bibinfo {year}
  {2012}{\natexlab{b}})}\BibitemShut {NoStop}%
\bibitem [{\citenamefont {Birowosuto}\ \emph
  {et~al.}(2014{\natexlab{b}})\citenamefont {Birowosuto}, \citenamefont
  {Yokoo}, \citenamefont {Zhang}, \citenamefont {Tateno}, \citenamefont
  {Kuramochi}, \citenamefont {Taniyama}, \citenamefont {Takiguchi},\ and\
  \citenamefont {Notomi}}]{Birowosuto2014}%
  \BibitemOpen
  \bibfield  {author} {\bibinfo {author} {\bibfnamefont {M.~D.}\ \bibnamefont
  {Birowosuto}}, \bibinfo {author} {\bibfnamefont {A.}~\bibnamefont {Yokoo}},
  \bibinfo {author} {\bibfnamefont {G.}~\bibnamefont {Zhang}}, \bibinfo
  {author} {\bibfnamefont {K.}~\bibnamefont {Tateno}}, \bibinfo {author}
  {\bibfnamefont {E.}~\bibnamefont {Kuramochi}}, \bibinfo {author}
  {\bibfnamefont {H.}~\bibnamefont {Taniyama}}, \bibinfo {author}
  {\bibfnamefont {M.}~\bibnamefont {Takiguchi}}, \ and\ \bibinfo {author}
  {\bibfnamefont {M.}~\bibnamefont {Notomi}},\ }\href@noop {} {\bibfield
  {journal} {\bibinfo  {journal} {Nature Mater.}\ }\textbf {\bibinfo {volume}
  {13}},\ \bibinfo {pages} {279} (\bibinfo {year}
  {2014}{\natexlab{b}})}\BibitemShut {NoStop}%
\bibitem [{\citenamefont {Chang}\ \emph {et~al.}(2006)\citenamefont {Chang},
  \citenamefont {S\o{}rensen}, \citenamefont {Hemmer},\ and\ \citenamefont
  {Lukin}}]{DEChang2006}%
  \BibitemOpen
  \bibfield  {author} {\bibinfo {author} {\bibfnamefont {D.~E.}\ \bibnamefont
  {Chang}}, \bibinfo {author} {\bibfnamefont {A.~S.}\ \bibnamefont
  {S\o{}rensen}}, \bibinfo {author} {\bibfnamefont {P.~R.}\ \bibnamefont
  {Hemmer}}, \ and\ \bibinfo {author} {\bibfnamefont {M.~D.}\ \bibnamefont
  {Lukin}},\ }\href@noop {} {\bibfield  {journal} {\bibinfo  {journal} {Phys.
  Rev. Lett.}\ }\textbf {\bibinfo {volume} {97}},\ \bibinfo {pages} {053002}
  (\bibinfo {year} {2006})}\BibitemShut {NoStop}%
\bibitem [{\citenamefont {Murai}\ \emph {et~al.}(2013)\citenamefont {Murai},
  \citenamefont {Verschuuren}, \citenamefont {Lozano}, \citenamefont
  {Pirruccio}, \citenamefont {Rodriguez},\ and\ \citenamefont
  {Rivas}}]{Murai2013}%
  \BibitemOpen
  \bibfield  {author} {\bibinfo {author} {\bibfnamefont {S.}~\bibnamefont
  {Murai}}, \bibinfo {author} {\bibfnamefont {M.~A.}\ \bibnamefont
  {Verschuuren}}, \bibinfo {author} {\bibfnamefont {G.}~\bibnamefont {Lozano}},
  \bibinfo {author} {\bibfnamefont {G.}~\bibnamefont {Pirruccio}}, \bibinfo
  {author} {\bibfnamefont {S.~R.~K.}\ \bibnamefont {Rodriguez}}, \ and\
  \bibinfo {author} {\bibfnamefont {J.~G.}\ \bibnamefont {Rivas}},\ }\href@noop
  {} {\bibfield  {journal} {\bibinfo  {journal} {Opt. Express}\ }\textbf
  {\bibinfo {volume} {21}},\ \bibinfo {pages} {4250} (\bibinfo {year}
  {2013})}\BibitemShut {NoStop}%
\bibitem [{\citenamefont {Akimov}\ \emph {et~al.}(2007)\citenamefont {Akimov},
  \citenamefont {Mukherjee}, \citenamefont {Yu}, \citenamefont {Chang},
  \citenamefont {Zibrov}, \citenamefont {Hemmer}, \citenamefont {Park},\ and\
  \citenamefont {Lukin}}]{Akimov2007}%
  \BibitemOpen
  \bibfield  {author} {\bibinfo {author} {\bibfnamefont {A.}~\bibnamefont
  {Akimov}}, \bibinfo {author} {\bibfnamefont {A.}~\bibnamefont {Mukherjee}},
  \bibinfo {author} {\bibfnamefont {C.}~\bibnamefont {Yu}}, \bibinfo {author}
  {\bibfnamefont {D.}~\bibnamefont {Chang}}, \bibinfo {author} {\bibfnamefont
  {A.}~\bibnamefont {Zibrov}}, \bibinfo {author} {\bibfnamefont
  {P.}~\bibnamefont {Hemmer}}, \bibinfo {author} {\bibfnamefont
  {H.}~\bibnamefont {Park}}, \ and\ \bibinfo {author} {\bibfnamefont
  {M.}~\bibnamefont {Lukin}},\ }\href@noop {} {\bibfield  {journal} {\bibinfo
  {journal} {Nature}\ }\textbf {\bibinfo {volume} {450}},\ \bibinfo {pages}
  {402} (\bibinfo {year} {2007})}\BibitemShut {NoStop}%
\bibitem [{\citenamefont {Kolesov}\ \emph {et~al.}(2009)\citenamefont
  {Kolesov}, \citenamefont {Grotz}, \citenamefont {Balasubramanian},
  \citenamefont {Stöhr}, \citenamefont {Nicolet}, \citenamefont {Hemmer},
  \citenamefont {Jelezko},\ and\ \citenamefont {Wrachtrup}}]{Kolesov2009}%
  \BibitemOpen
  \bibfield  {author} {\bibinfo {author} {\bibfnamefont {R.}~\bibnamefont
  {Kolesov}}, \bibinfo {author} {\bibfnamefont {B.}~\bibnamefont {Grotz}},
  \bibinfo {author} {\bibfnamefont {G.}~\bibnamefont {Balasubramanian}},
  \bibinfo {author} {\bibfnamefont {R.}~\bibnamefont {Stöhr}}, \bibinfo
  {author} {\bibfnamefont {A.}~\bibnamefont {Nicolet}}, \bibinfo {author}
  {\bibfnamefont {P.}~\bibnamefont {Hemmer}}, \bibinfo {author} {\bibfnamefont
  {F.}~\bibnamefont {Jelezko}}, \ and\ \bibinfo {author} {\bibfnamefont
  {J.}~\bibnamefont {Wrachtrup}},\ }\href@noop {} {\bibfield  {journal}
  {\bibinfo  {journal} {Nature Physics}\ }\textbf {\bibinfo {volume} {5}},\
  \bibinfo {pages} {470} (\bibinfo {year} {2009})}\BibitemShut {NoStop}%
\bibitem [{\citenamefont {Huck}\ \emph {et~al.}(2011)\citenamefont {Huck},
  \citenamefont {Kumar}, \citenamefont {Shakoor},\ and\ \citenamefont
  {Andersen}}]{AlexHuck2011}%
  \BibitemOpen
  \bibfield  {author} {\bibinfo {author} {\bibfnamefont {A.}~\bibnamefont
  {Huck}}, \bibinfo {author} {\bibfnamefont {S.}~\bibnamefont {Kumar}},
  \bibinfo {author} {\bibfnamefont {A.}~\bibnamefont {Shakoor}}, \ and\
  \bibinfo {author} {\bibfnamefont {U.~L.}\ \bibnamefont {Andersen}},\
  }\href@noop {} {\bibfield  {journal} {\bibinfo  {journal} {Phys. Rev. Lett.}\
  }\textbf {\bibinfo {volume} {106}},\ \bibinfo {pages} {096801} (\bibinfo
  {year} {2011})}\BibitemShut {NoStop}%
\bibitem [{\citenamefont {Frimmer}, \citenamefont {Chen},\ and\ \citenamefont
  {Koenderink}(2011)}]{Frimmer2011}%
  \BibitemOpen
  \bibfield  {author} {\bibinfo {author} {\bibfnamefont {M.}~\bibnamefont
  {Frimmer}}, \bibinfo {author} {\bibfnamefont {Y.}~\bibnamefont {Chen}}, \
  and\ \bibinfo {author} {\bibfnamefont {A.~F.}\ \bibnamefont {Koenderink}},\
  }\href@noop {} {\bibfield  {journal} {\bibinfo  {journal} {Phys. Rev. Lett.}\
  }\textbf {\bibinfo {volume} {107}},\ \bibinfo {pages} {123602} (\bibinfo
  {year} {2011})}\BibitemShut {NoStop}%
\bibitem [{\citenamefont {Pyayt}\ \emph {et~al.}(2008)\citenamefont {Pyayt},
  \citenamefont {Wiley}, \citenamefont {Xia}, \citenamefont {Chen},\ and\
  \citenamefont {Dalton}}]{Pyayt2008}%
  \BibitemOpen
  \bibfield  {author} {\bibinfo {author} {\bibfnamefont {A.}~\bibnamefont
  {Pyayt}}, \bibinfo {author} {\bibfnamefont {B.}~\bibnamefont {Wiley}},
  \bibinfo {author} {\bibfnamefont {Y.}~\bibnamefont {Xia}}, \bibinfo {author}
  {\bibfnamefont {A.}~\bibnamefont {Chen}}, \ and\ \bibinfo {author}
  {\bibfnamefont {L.}~\bibnamefont {Dalton}},\ }\href@noop {} {\bibfield
  {journal} {\bibinfo  {journal} {Nature Nanotechnology}\ }\textbf {\bibinfo
  {volume} {3}},\ \bibinfo {pages} {660} (\bibinfo {year} {2008})}\BibitemShut
  {NoStop}%
\bibitem [{\citenamefont {Kumar}\ \emph {et~al.}(2013)\citenamefont {Kumar},
  \citenamefont {Huck}, \citenamefont {Chen},\ and\ \citenamefont
  {Andersen}}]{AlexHuck2013a}%
  \BibitemOpen
  \bibfield  {author} {\bibinfo {author} {\bibfnamefont {S.}~\bibnamefont
  {Kumar}}, \bibinfo {author} {\bibfnamefont {A.}~\bibnamefont {Huck}},
  \bibinfo {author} {\bibfnamefont {Y.}~\bibnamefont {Chen}}, \ and\ \bibinfo
  {author} {\bibfnamefont {U.~L.}\ \bibnamefont {Andersen}},\ }\href@noop {}
  {\bibfield  {journal} {\bibinfo  {journal} {Appl. Phys. Lett.}\ }\textbf
  {\bibinfo {volume} {102}},\ \bibinfo {pages} {103106} (\bibinfo {year}
  {2013})}\BibitemShut {NoStop}%
\bibitem [{\citenamefont {Kumar}, \citenamefont {Huck},\ and\ \citenamefont
  {Andersen}(2013)}]{AlexHuck2013}%
  \BibitemOpen
  \bibfield  {author} {\bibinfo {author} {\bibfnamefont {S.}~\bibnamefont
  {Kumar}}, \bibinfo {author} {\bibfnamefont {A.}~\bibnamefont {Huck}}, \ and\
  \bibinfo {author} {\bibfnamefont {U.~L.}\ \bibnamefont {Andersen}},\
  }\href@noop {} {\bibfield  {journal} {\bibinfo  {journal} {Nano Lett.}\
  }\textbf {\bibinfo {volume} {13}},\ \bibinfo {pages} {1221} (\bibinfo {year}
  {2013})}\BibitemShut {NoStop}%
\bibitem [{\citenamefont {Chance}, \citenamefont {Prock},\ and\ \citenamefont
  {Silbey}(1978)}]{CPS1978}%
  \BibitemOpen
  \bibfield  {author} {\bibinfo {author} {\bibfnamefont {R.~R.}\ \bibnamefont
  {Chance}}, \bibinfo {author} {\bibfnamefont {A.}~\bibnamefont {Prock}}, \
  and\ \bibinfo {author} {\bibfnamefont {R.}~\bibnamefont {Silbey}},\ }\href
  {\doibase 10.1002/9780470142561.ch1} {\bibfield  {journal} {\bibinfo
  {journal} {Adv. in Chem. Phys.}\ ,\ \bibinfo {pages} {1}} (\bibinfo {year}
  {1978})}\BibitemShut {NoStop}%
\bibitem [{\citenamefont {Amos}\ and\ \citenamefont {Barnes}(1997)}]{Amos1997}%
  \BibitemOpen
  \bibfield  {author} {\bibinfo {author} {\bibfnamefont {R.~M.}\ \bibnamefont
  {Amos}}\ and\ \bibinfo {author} {\bibfnamefont {W.~L.}\ \bibnamefont
  {Barnes}},\ }\href@noop {} {\bibfield  {journal} {\bibinfo  {journal} {Phys.
  Rev. B}\ }\textbf {\bibinfo {volume} {55}},\ \bibinfo {pages} {7249}
  (\bibinfo {year} {1997})}\BibitemShut {NoStop}%
\bibitem [{\citenamefont {Snoeks}, \citenamefont {Lagendijk},\ and\
  \citenamefont {Polman}(1995)}]{Snoeks1995}%
  \BibitemOpen
  \bibfield  {author} {\bibinfo {author} {\bibfnamefont {E.}~\bibnamefont
  {Snoeks}}, \bibinfo {author} {\bibfnamefont {A.}~\bibnamefont {Lagendijk}}, \
  and\ \bibinfo {author} {\bibfnamefont {A.}~\bibnamefont {Polman}},\
  }\href@noop {} {\bibfield  {journal} {\bibinfo  {journal} {Phys. Rev. Lett.}\
  }\textbf {\bibinfo {volume} {74}},\ \bibinfo {pages} {2459} (\bibinfo {year}
  {1995})}\BibitemShut {NoStop}%
\bibitem [{\citenamefont {Kippenberg}\ \emph {et~al.}(2006)\citenamefont
  {Kippenberg}, \citenamefont {Kalkman}, \citenamefont {Polman},\ and\
  \citenamefont {Vahala}}]{Kippenberg2006}%
  \BibitemOpen
  \bibfield  {author} {\bibinfo {author} {\bibfnamefont {T.~J.}\ \bibnamefont
  {Kippenberg}}, \bibinfo {author} {\bibfnamefont {J.}~\bibnamefont {Kalkman}},
  \bibinfo {author} {\bibfnamefont {A.}~\bibnamefont {Polman}}, \ and\ \bibinfo
  {author} {\bibfnamefont {K.~J.}\ \bibnamefont {Vahala}},\ }\href@noop {}
  {\bibfield  {journal} {\bibinfo  {journal} {Phys. Rev. A}\ }\textbf {\bibinfo
  {volume} {74}},\ \bibinfo {pages} {051802} (\bibinfo {year}
  {2006})}\BibitemShut {NoStop}%
\bibitem [{\citenamefont {Kroekenstoel}\ \emph {et~al.}(2009)\citenamefont
  {Kroekenstoel}, \citenamefont {Verhagen}, \citenamefont {Walters},
  \citenamefont {Kuipers},\ and\ \citenamefont {Polman}}]{Kroekenstoel2009}%
  \BibitemOpen
  \bibfield  {author} {\bibinfo {author} {\bibfnamefont {E.~J.~A.}\
  \bibnamefont {Kroekenstoel}}, \bibinfo {author} {\bibfnamefont
  {E.}~\bibnamefont {Verhagen}}, \bibinfo {author} {\bibfnamefont {R.~J.}\
  \bibnamefont {Walters}}, \bibinfo {author} {\bibfnamefont {L.}~\bibnamefont
  {Kuipers}}, \ and\ \bibinfo {author} {\bibfnamefont {A.}~\bibnamefont
  {Polman}},\ }\href@noop {} {\bibfield  {journal} {\bibinfo  {journal} {Appl.
  Phys. Lett.}\ }\textbf {\bibinfo {volume} {95}},\ \bibinfo {eid} {263106}
  (\bibinfo {year} {2009})}\BibitemShut {NoStop}%
\bibitem [{\citenamefont {Chew}(1988)}]{Chew1988}%
  \BibitemOpen
  \bibfield  {author} {\bibinfo {author} {\bibfnamefont {H.}~\bibnamefont
  {Chew}},\ }\href@noop {} {\bibfield  {journal} {\bibinfo  {journal} {Phys.
  Rev. A}\ }\textbf {\bibinfo {volume} {38}},\ \bibinfo {pages} {3410}
  (\bibinfo {year} {1988})}\BibitemShut {NoStop}%
\bibitem [{\citenamefont {Hohenester}\ and\ \citenamefont
  {Trügler}(2012)}]{Hohenester2012}%
  \BibitemOpen
  \bibfield  {author} {\bibinfo {author} {\bibfnamefont {U.}~\bibnamefont
  {Hohenester}}\ and\ \bibinfo {author} {\bibfnamefont {A.}~\bibnamefont
  {Trügler}},\ }\href@noop {} {\bibfield  {journal} {\bibinfo  {journal}
  {Comp. Phys. Comm.}\ }\textbf {\bibinfo {volume} {183}},\ \bibinfo {pages}
  {370 } (\bibinfo {year} {2012})}\BibitemShut {NoStop}%
\bibitem [{\citenamefont {Birowosuto}\ \emph {et~al.}(2010)\citenamefont
  {Birowosuto}, \citenamefont {Skipetrov}, \citenamefont {Vos},\ and\
  \citenamefont {Mosk}}]{Birowosuto2010}%
  \BibitemOpen
  \bibfield  {author} {\bibinfo {author} {\bibfnamefont {M.~D.}\ \bibnamefont
  {Birowosuto}}, \bibinfo {author} {\bibfnamefont {S.~E.}\ \bibnamefont
  {Skipetrov}}, \bibinfo {author} {\bibfnamefont {W.~L.}\ \bibnamefont {Vos}},
  \ and\ \bibinfo {author} {\bibfnamefont {A.~P.}\ \bibnamefont {Mosk}},\
  }\href@noop {} {\bibfield  {journal} {\bibinfo  {journal} {Phys. Rev. Lett.}\
  }\textbf {\bibinfo {volume} {105}},\ \bibinfo {pages} {013904} (\bibinfo
  {year} {2010})}\BibitemShut {NoStop}%
\bibitem [{\citenamefont {Cesa}\ \emph {et~al.}(2009)\citenamefont {Cesa},
  \citenamefont {Blum}, \citenamefont {van~den Broek}, \citenamefont {Mosk},
  \citenamefont {Vos},\ and\ \citenamefont {Subramaniam}}]{Cesa2009}%
  \BibitemOpen
  \bibfield  {author} {\bibinfo {author} {\bibfnamefont {Y.}~\bibnamefont
  {Cesa}}, \bibinfo {author} {\bibfnamefont {C.}~\bibnamefont {Blum}}, \bibinfo
  {author} {\bibfnamefont {J.~M.}\ \bibnamefont {van~den Broek}}, \bibinfo
  {author} {\bibfnamefont {A.~P.}\ \bibnamefont {Mosk}}, \bibinfo {author}
  {\bibfnamefont {W.~L.}\ \bibnamefont {Vos}}, \ and\ \bibinfo {author}
  {\bibfnamefont {V.}~\bibnamefont {Subramaniam}},\ }\href@noop {} {\bibfield
  {journal} {\bibinfo  {journal} {Phys. Chem. Chem. Phys.}\ }\textbf {\bibinfo
  {volume} {11}},\ \bibinfo {pages} {2525} (\bibinfo {year}
  {2009})}\BibitemShut {NoStop}%
\bibitem [{\citenamefont {Garc\'{i}a~de Abajo}\ and\ \citenamefont
  {Howie}(2002)}]{FJAbajo2002}%
  \BibitemOpen
  \bibfield  {author} {\bibinfo {author} {\bibfnamefont {F.~J.}\ \bibnamefont
  {Garc\'{i}a~de Abajo}}\ and\ \bibinfo {author} {\bibfnamefont
  {A.}~\bibnamefont {Howie}},\ }\href@noop {} {\bibfield  {journal} {\bibinfo
  {journal} {Phys. Rev. B}\ }\textbf {\bibinfo {volume} {65}},\ \bibinfo
  {pages} {115418} (\bibinfo {year} {2002})}\BibitemShut {NoStop}%
\end{thebibliography}
\end{document}